\documentclass[aps,prd,reprint,nofootinbib]{revtex4-2}
\DeclareUnicodeCharacter{2009}{\,}
\usepackage{xspace}
\usepackage{amsmath}
\usepackage{graphicx}
\usepackage[justification=justified, singlelinecheck=off]{caption}
\usepackage{csvsimple}
\usepackage{booktabs}
\usepackage{nicematrix}
\usepackage{orcidlink}
\usepackage{subcaption}
\graphicspath{{./figures/}}
\usepackage{ragged2e}
\usepackage[normalem]{ulem}
\usepackage[utf8]{inputenc}
\usepackage{todonotes}

\begin{document}
\preprint{}

\title{Is Gravitational-Wave Waveform Systematics a Bottleneck for Formation Channel Inferences using Effective-Spin and Mass-Ratio?}
\author{Upasana Das \orcidlink{0009-0009-0307-2265}}
\email{upasana.das@ligo.org} \email{upasana806@gmail.com}
\affiliation{%
Institute of Astronomy, KU Leuven, Celestijnenlaan 200D, 3001 Leuven, Belgium 
}%

\author{Suvodip Mukherjee \orcidlink{0000-0002-3373-5236}}%
 \email{suvodip@tifr.res.in }
\affiliation{%
Department of Astronomy and Astrophysics, Tata Institute of
Fundamental Research, 1, Homi Bhabha Road, Mumbai-400005,
Maharashtra, India
}%



\begin{abstract}
Gravitational-wave measurements of the effective spin $\chi_\text{eff}$ and mass ratio $q$ help distinguish binary black hole formation channels, motivating a search for structure in the $\chi_\text{eff} - q$ plane. However, both these quantities are known to be strongly correlated in GW parameter estimation since both enter the leading-order spin-orbit term together, and thus, the detectors cannot easily tell them apart. In this work, we harness the susceptibility of such measurements to waveform systematics by using injection-recovery studies with \texttt{NRSur7dq4} as the true signal and \texttt{SEOBNRv5PHM} and \texttt{IMRPhenomTPHM} for parameter inferences. Our study shows that, though there can be some parameter space in the $\chi_\text{eff} -  q$ plane, primarily in the unequal-mass regime, there can be some impacts from waveform systematics; overall, at a population level, the waveform systematics for the current detectors at their design sensitivity are not a bottleneck in recovering the underlying true population properties. We find that astrophysical properties such as $\chi_\text{eff} - q$ correlations, which are signatures of different formation channels, are unlikely to be spuriously induced by waveform systematics, and inferences about formation channels based on such correlations are likely robust to waveform systematics at present, though some events that are in the parameter space of large impact can show some outlier behaviour. We find that some of the GW events such as GW190403$\_$051519 and GW240527$\_$230910 are in the vicinity of the $\chi_\text{eff} -  q$ parameter space, where large waveform systematics can arise, and these events are more susceptible to waveform systematics than others in GWTC-5. This study will play a vital role in precision astrophysical population inference from GW data in the future and in assessing the possibility of waveform systematics.  
\end{abstract}


\maketitle
\section{Introduction}%
\label{sec:Introduction}

The direct detection of gravitational waves (GWs) from the merger of two stellar-mass black holes in 2015 \cite{Abbott1} opened a new observational channel for studying compact objects and demonstrated that the observed waveform can be used to infer the physical properties of the source. Since then, the LIGO-Virgo-KAGRA (LVK) detectors \cite{LIGO, LVK1, Virgo2, Kagra} have accumulated hundreds of such compact-binary candidates \cite{GWTC5_intro}, making it possible to move beyond the study of individual events to population-level analyses, which combine the full catalog to infer the statistical distribution of source properties across the underlying binary black hole (BBH) population. The robustness of such a study to the choice of waveform model used in parameter estimation (PE) is what motivates our present work.

Among the source properties accessible from the GW signal, the spins of the component black holes are especially difficult to measure. At the order at which spin first enters the post-Newtonian (PN) phase, 1.5PN, via spin-orbit coupling, the waveform depends only on a single effective spin parameter $\chi_\text{eff}$, quantifying the net projection of both component spins onto the binary’s orbital angular momentum \cite{Kidder1995, PoissonWill1995, Flanagan1994}.
\begin{equation}
\chi_{\rm eff} \equiv \frac{m_1\vec{\chi}_1\cdot\hat{L} + m_2\vec{\chi}_2\cdot\hat{L}}{m_1+m_2},
\label{eq:chieff}
\end{equation}
where $m_1\geq m_2$ are the component masses, $\hat{L}$ is the unit vector along the orbital angular momentum and $\vec{\chi}_{1,2}$ are the dimensionless component spins. Because $\chi_{\rm eff}$ remains nearly conserved during the inspiral \cite{Racine08, Gerosa2015}, it is measured with precision from the GW phase and, together with the mass ratio, $q\equiv m_2/m_1$, is also the source parameter most directly connected to a binary's formation history. 

 Different formation channels can leave characteristic signatures in this $(q, \chi_{\rm eff})$ plane \cite{Vitale_2017}. In the isolated binary-evolution channel, the common-envelope pathway is expected to produce highly aligned, low-to-moderate spins, although tidal spin-up can generate a positive-$\chi_{\rm eff}$ tail \cite{BaveraCE, Zevin_2021}. Chemically homogeneous evolution, another isolated channel, preferentially produces nearly equal-mass BBHs with aligned spins and can therefore populate the high-$\chi_{\rm eff}, q \simeq 1$ region \cite{Mandel_2016}. Dynamically assembled binaries in dense stellar environments are theorised to acquire isotropic spin orientations through repeated interactions \cite{dynamic, dynamic2}, giving a broad distribution centered at zero. BBHs formed in active galactic nucleus (AGN) disks can produce distinct localised, environment-driven correlations between $q$ and $\chi_{\rm eff}$ \cite{AGN_channel}. Hierarchical mergers are an additional pathway in which merger remnants with characteristic spin magnitudes of $\sim 0.7$ participate in subsequent mergers, potentially producing high-spin systems and, for 2G+1G mergers, asymmetric mass ratios around $q \sim 0.5$ \cite{Doctor_2020}.



These expectations also motivate studying the joint distribution of $q$ and $\chi_{\rm eff}$. An intrinsic anti-correlation between these two parameters has been observed in \cite{Callister_2021}, where they found that unequal-mass BBHs tend to have larger  $\chi_{\rm eff}$ than equal-mass BBHs, with $98.7\%$ credibility for a correlation between the two parameters. Proposed astrophysical explanations for this correlation include differences in the relative contributions of isolated and dynamically assembled systems, as well as formation in AGN disks \cite{McKernan22}. The statistical description of this correlation has also been examined using alternative population models \cite{Adamcewicz_2022}. The latest population analyses continue to find structure in these parameters. GWTC-4.0 population analysis found that the joint $(q,\chi_{\rm eff})$ distribution transitions in shape across distinct primary-mass regions, with some regions showing markedly reduced support for near-equal mass recoveries \cite{Region4}, and updated population fits using GWTC-5.0 further resolve this  \cite{GWTC5, Transitions2026, Antonini2026}. These results illustrate how features in this plane are increasingly being used to identify distinct BBH formation channels.


Such interpretations, however, rest on the assumption that the inferred structure is entirely astrophysical rather than considering waveform systematics, an assumption that is easy to make and difficult to check since $q$ and $\chi_{\rm eff}$ are measured degenerately \cite{degeneracyqchi}. The mass ratio enters the phase at leading order through the symmetric mass ratio $\eta \equiv q/(1+q)^2$, while the spin-orbit coupling term, introduced at 1.5PN order, is degenerate with $\eta$. Consequently, changes in the observed phase evolution can be partially compensated by changes in both $\eta$ and $\chi_{\rm eff}$, producing a degeneracy between these quantities in event-level PE. This raises the question: could some of the features in the observed $\chi_{\rm eff}-q$ observations be generated by waveform systematics, especially of the kind that are currently being interpreted as physical evidence for distinct formation channels? A parallel line of work has also recently raised waveform systematics as one possible source of population-level correlations. \cite{Similarwork}.


In a companion study (hereafter Paper-I) \cite{Paper1}, we showed that waveform systematics can induce mass-dependent shifts in the recovery of component masses and that these event-level shifts propagate through hierarchical population inference to distort features in the recovered BBH mass spectrum. Here we take this one step further and ask how the same systematics can propagate from any biased event-level measurements into the population-level analyses used to interpret BBH formation channels.




The remainder of this paper is organised as follows. Section \ref{sec:Methods} briefly describes the generation of the mock catalog (details in Appendix \ref{app:Setup deets}), the injection-recovery setup and the definitions of our shift in parameters. Section \ref{sec:Results} reports the event-level shifts for the two parameters, first, individually (Sec. \ref{subsec:individual}), then jointly (Sec. \ref{subsec:joint}). We go on to contextualise these results by overlaying GWTC-5 events onto our simulated parameter space (Sec. \ref{subsec:gwtc5}). Finally, Section \ref{sec:Disc_and_concl} discusses the implications of these findings for population-level analyses, along with future scope of this study and concludes the paper.

\section{Methods}
\label{sec:Methods}

\subsection{Mock Population catalog}
We construct a mock BBH catalog using \texttt{GWSim}\cite{GWSim}, following the same population and detector setup as in Paper-I.  The BBH merger rate density follows the Madau-Dickinson star formation history \cite{MadauDickinson}, with a local merger rate of $R_0 = 15\, \text{Gpc}^{-3} \text{yr}^{-1}$ and an observation time of one year, assuming a $\Lambda$CDM cosmology consistent with Planck-2018 results \cite{Planck18}. 

\begin{figure}[htbp]
    \centering
    \includegraphics[width=1.05\columnwidth]{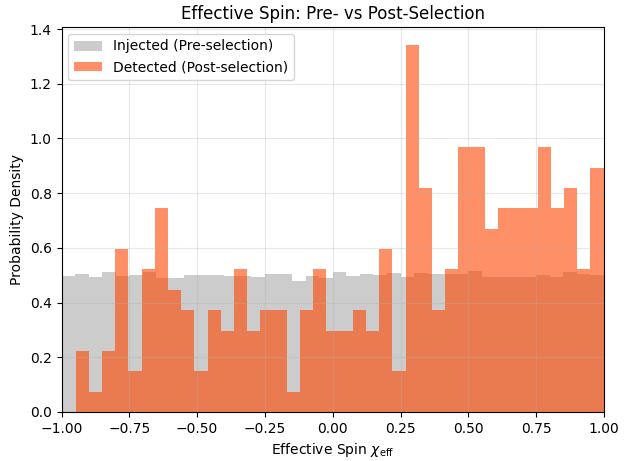}
    
    \includegraphics[width=1.05\columnwidth]{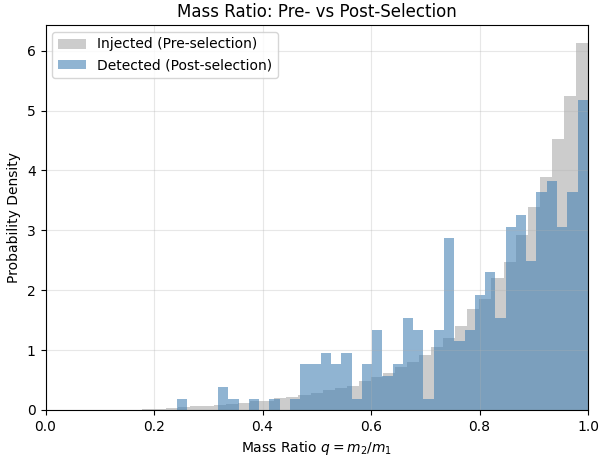}
    
    \caption{\justifying Pre- (grey) vs post-selection (coloured) probability distributions for $\chi_{\rm eff}$ (top) and $q$ (bottom) of our mock catalog.} 
    \label{fig:pre-vs-post}
\end{figure}

The primary mass $m_1$ is drawn from the same Power Law+Gaussian (PL+G) distribution, again the same as in Paper-I, and the secondary mass $m_2$ follows a power law distribution such that $m_2 \leq m_1$. The full population model and hyperparameters are given in Appendix \ref{app:Setup deets}. The effective spin $\chi_\text{eff}$ is drawn independently of mass from a uniform distribution over $[-1,1]$, corresponding to a population with no preferred spin orientation and no intrinsic mass-spin correlation by construction. We note that a uniformly generated $\chi_\text{eff}$ need not remain uniform after selection since aligned-spin systems are intrinsically louder at a fixed mass \cite{orbital_hangup, Ng}. We show the pre- vs post-selection $\chi_{\rm eff}$ and $q$ distributions in Fig. \ref{fig:pre-vs-post}.  Detector selection effects are incorporated by computing the matched-filter signal-to-noise ratio (SNR) for each injected signal, assuming the O5 noise power spectral density (PSD) for the H1-L1 network \cite{MatchedFilter, PSD,psd2,psd3}. We retain events with matched-filtering SNR above a threshold of $12$, resulting in a detected sample of $N=276$ events that constitutes our injection catalog.  

\subsection{Injection-Recovery setup}
For each event in the mock catalog, the injected signal is constructed using the \texttt{NRSur7dq4} surrogate model \cite{NRSur_foundational, NRSur, NRSur_new}, adopted here as a benchmark model rather than as ground truth. Each event is then recovered using \texttt{SEOBNRv5PHM} \cite{SEOB_foundational, SEOB1, SEOB2} and \texttt{IMRPhenomTPHM} \cite{IMRPhenom1, IMRPhenom2, TPHM}, with PE performed in Bilby \cite{Bilby, Bilby2, bilby_paper} using the Nessai \cite{nessai, nessai2, nessai3} sampler, with sampler settings and a fixed noise seed identical to Paper-I. The priors on the recovered parameters and other details are listed in Appendix \ref{app:Setup deets}.

\subsection{Shift Definitions}
\label{subsec:biasdefn}
For each event and recovery waveform, we define the systematic shift in the effective spin and mass ratio as the difference between the posterior median and the injected value.

\begin{align}
\Delta\chi_{\rm eff} &\equiv \mathrm{median}(\chi_{\rm eff,rec}) - \chi_{\rm eff,true}, \\
\Delta q &\equiv \mathrm{median}(q_{\rm rec}) - q_{\rm true}.
\label{eq:bias_def}
\end{align}

To quantify regions of the $(q_{\rm true},\chi_{\rm eff,true})$ plane where the recovery produces the largest combined displacement, we define a per-event joint metric in unweighted ($\Delta^2_{\rm raw}$) and weighted  ($\Delta^2_{\sigma}$)  forms:

\begin{align}
\Delta^2_{\rm raw} &\equiv (\Delta\chi_{\rm eff})^2 + (\Delta q)^2, \label{eq:metric_raw}\\
\Delta^2_{\sigma} &\equiv \left(\frac{\Delta\chi_{\rm eff}}{\sigma_{\chi_{\rm eff}}}\right)^2 + \left(\frac{\Delta q}{\sigma_q}\right)^2, \label{eq:metric_weighted}
\end{align}
where $\sigma_{\chi_{\rm eff}}$ and $\sigma_q$ are the $68\%$ credible interval  widths of the corresponding posteriors. $\Delta_{\rm raw}$ measures the absolute displacement from the injected parameters, $\Delta_\sigma$ identifies where this displacement is most significant relative to each event's own measurement uncertainty, and hence most likely to distort population-level inference. 


\section{Results}
\label{sec:Results}
We present the results of the PE study performed on the injected catalog of events described in Sec. \ref{sec:Methods}. We first examine the individual-event shift in $\chi_{\rm eff}$ and $q$ separately, then their joint behaviour, and finally compare the results recovered space against the real GWTC-5.0 catalog, to check whether the  GW events overlap with the parameter space impacted by mismatch between injected value and the recovered value.



\subsection{$\chi_\text{eff}$ and $q$ recovery: Individual Event}
\label{subsec:individual}

\begin{figure*}[htbp] 
    \centering
    \begin{minipage}[t]{0.5\textwidth} 
        \centering
        \includegraphics[width=\linewidth]{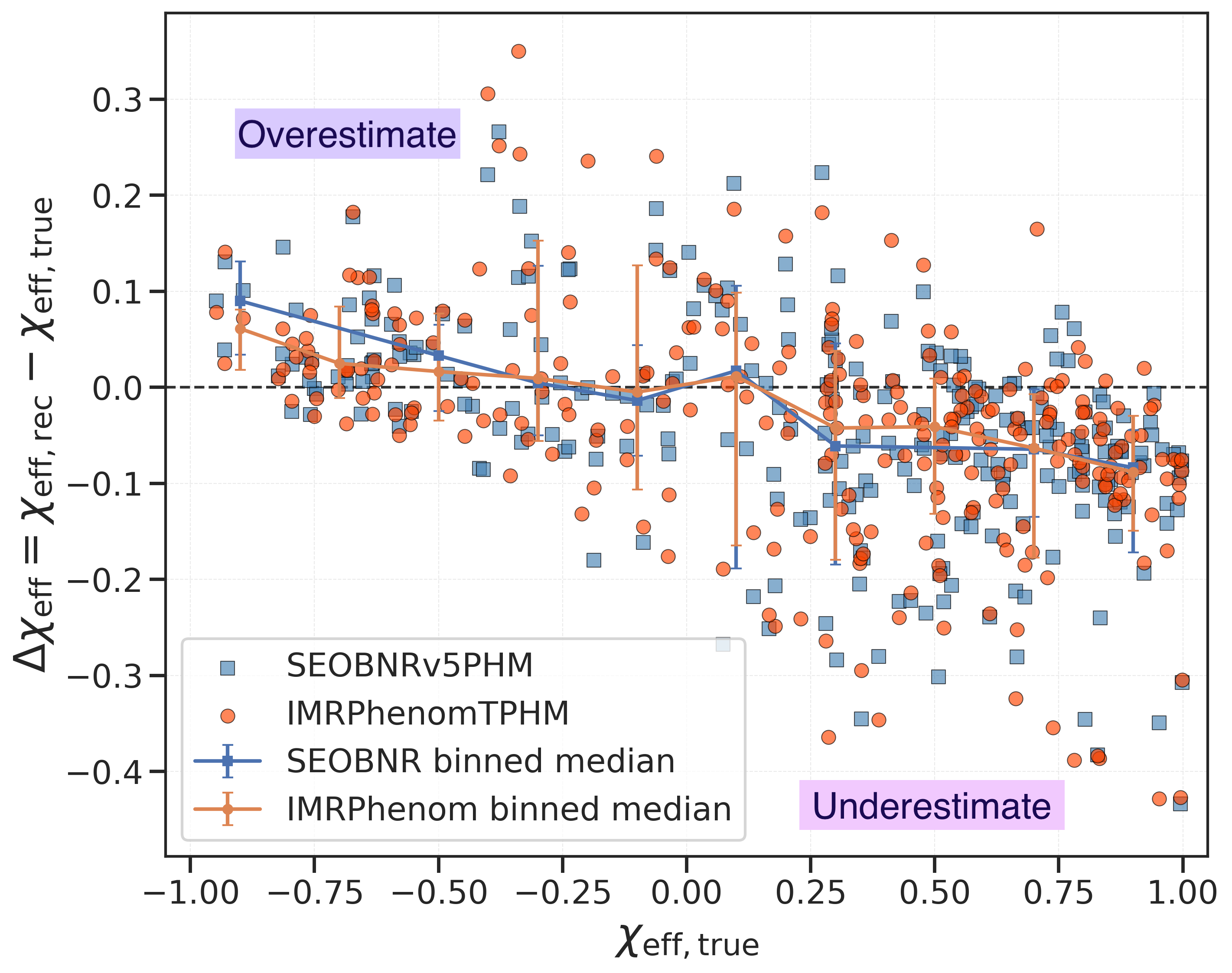} 
        \label{fig:figure1}
    \end{minipage}%
    \hfill 
    \begin{minipage}[t]{0.5\textwidth} 
        \centering
        \includegraphics[width=\linewidth]{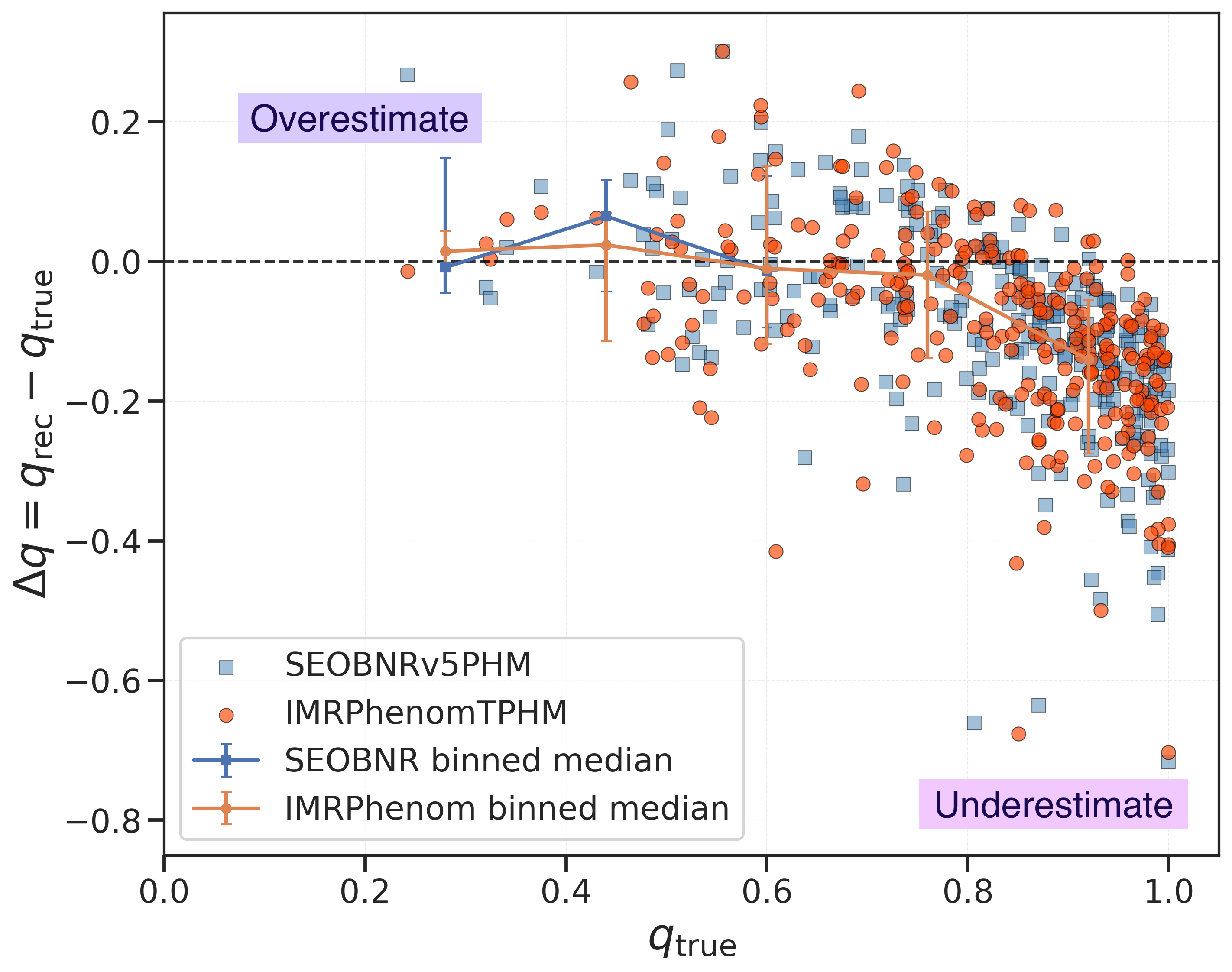}
        \label{fig:figure2}
    \end{minipage}
    \caption{\justifying \justifying{ Left: Systematic shift in the recovered effective spin, $\Delta \chi_\text{eff} = \chi_\text{eff,rec} - \chi_\text{eff,true}$ vs the true injected $\chi_\text{eff, true}$. Right: Systematic shift in the recovered mass ratio $\Delta q = q_\text{rec} - q_\text{true}$ vs the true injected $q_{\rm true}$. Both panels show events recovered with \texttt{SEOBNRv5PHM} (blue squares) and \texttt{IMRPhenomTPHM} (red circles) against an \texttt{NRSur7dq4} injection. Solid lines show the median shift in bins of the true value, with error bars spanning the 16th-84th percentile spread of individual-event shift within each bin. The dashed line corresponds to unbiased recovery.}} 
    \label{fig:individual_bias}
\end{figure*}

\begin{figure*}[htbp]
    \centering
  \includegraphics[width=\linewidth]{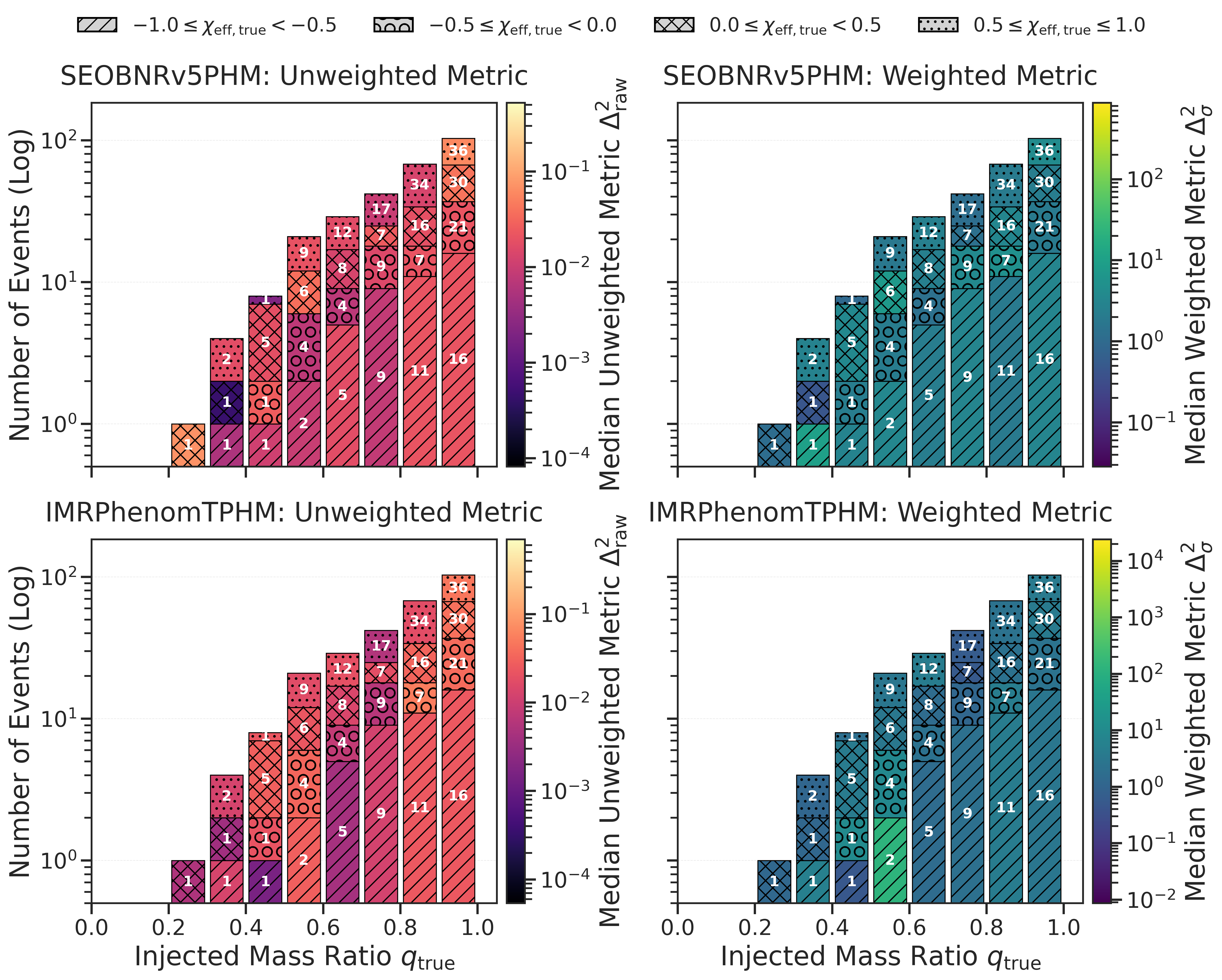}   \caption{\justifying{Injected events binned by $q_{\rm true}$. Within each bin, events are further split by $\chi_{\rm eff, true}$ and coloured by the median value of the unweighted metric $\Delta^2_{\rm raw}$ (left) or the weighted metric $\Delta^2_{\sigma}$ (right)}}
    \label{fig:metric_colour}
\end{figure*}

The left panel of Fig.  \ref{fig:individual_bias} shows the systematic shift in the recovered effective spin $\Delta\chi_{\rm eff}$ as a function of the injected value. Both waveform recoveries show a similar dependence of the shift on $\chi_{\rm eff, true}$: strongly anti-aligned systems with negative $\chi_{\rm eff, true}$ are preferentially recovered at less negative values (positive $\Delta\chi_{\rm eff}$), whereas aligned-spin systems with positive $\chi_{\rm eff,true}$ are shifted towards smaller effective spins (negative $\Delta\chi_{\rm eff}$), developing an anti-correlation between any shift in the recovered values and injected parameters. Correspondingly, the top-right quadrant (large positive true value, large positive shift) and bottom-left quadrant (large negative true value, large negative shift) are both sparsely populated: recovery essentially never pushes an already-extreme spin further from zero. The overall effect is a compression of the recovered $\chi_{\rm eff}$ distribution towards intermediate values, with the shift being comparatively small but with large individual-event scatter near $\chi_\text{eff,true} \approx 0.0$.  This compression directly affects the credibility of the very observable that primarily distinguishes between different formation channels. The right panel of Fig. \ref{fig:individual_bias} shows the equivalent systematic shift in the recovered mass ratio $\Delta q$. Compared to the more scattered $\chi_{\rm eff}$ behaviour, the shift in $q$ follows a more monotonic dependence. Recovery is mildly positive for the most asymmetric systems ($q_{\rm true}\lesssim0.4$), scattered around zero for $q_{\rm true}\sim0.4$--$0.5$, and becomes increasingly negative as $q_{\rm true}\to 1$. Near-equal-mass systems (and absolute value of $\chi_{\rm eff}$) close to one are preferentially recovered away from the $\Delta q=0$ (and $\Delta \chi_{\rm eff}= 0$ line), which is primarily due to prior effect. For events that are close to the boundary values of the parameters, the posterior distribution of the recovered values rails against the prior boundary; as a result, the median value is always shifted. We return to this later in Sec \ref{subsec:gwtc5}. The sparsity of injected events below $q_{\rm true} \sim 0.3$ is set by the population model because the steep primary-mass power law ($\alpha = 3.5$) concentrates $m_1$ near the minimum mass, and since $m_2 \ge m_{\rm min} = 5 M_\odot$, small mass ratios are rare even before selection.




We examine each event's displacement in the $(q,\chi_{\rm eff})$ plane using the unweighted and weighted metrics defined in Sec. \ref{subsec:biasdefn}. Figure \ref{fig:metric_colour} bins the injected catalog by $q_{\rm true}$ and within each mass ratio bin, further splits events by $\chi_{\rm eff, true}$, coloured by the median value of $\Delta^2_{\rm raw}$ (left) and $\Delta^2_{\sigma}$ (right) within each sub-bin. This lets us ask, directly, where in the two-dimensional $(q,\chi_{\rm eff})$ parameter space the joint recovery is displaced the most in absolute terms (via $\Delta^2_{\rm raw}$), and where that displacement is most statistically significant relative to each event's own measurement uncertainty (via $\Delta^2_{\sigma}$) and hence most likely to affect the inferred population structure. For both waveform families, the median  $\Delta^2_{\rm raw}$ increases systematically toward lower $q_{\rm true}$, with variation across the $\chi_{\rm eff, true}$ bins, and the $\Delta^2_{\sigma}$ shows a similar trend reflecting that measurement precision itself also varies across this plane. We also identify exact pockets of parameter space where recovery is worst that will help analyses be more careful where events might be detected. Although, note that the number of events varies substantially across bins, so medians in sparsely populated regions should be interpreted cautiously.

\subsection{Joint $\chi_\text{eff}-q$ shift: Population level}
\label{subsec:joint}

\begin{figure*}[htbp]
    \centering
   \includegraphics[width=0.8\linewidth]{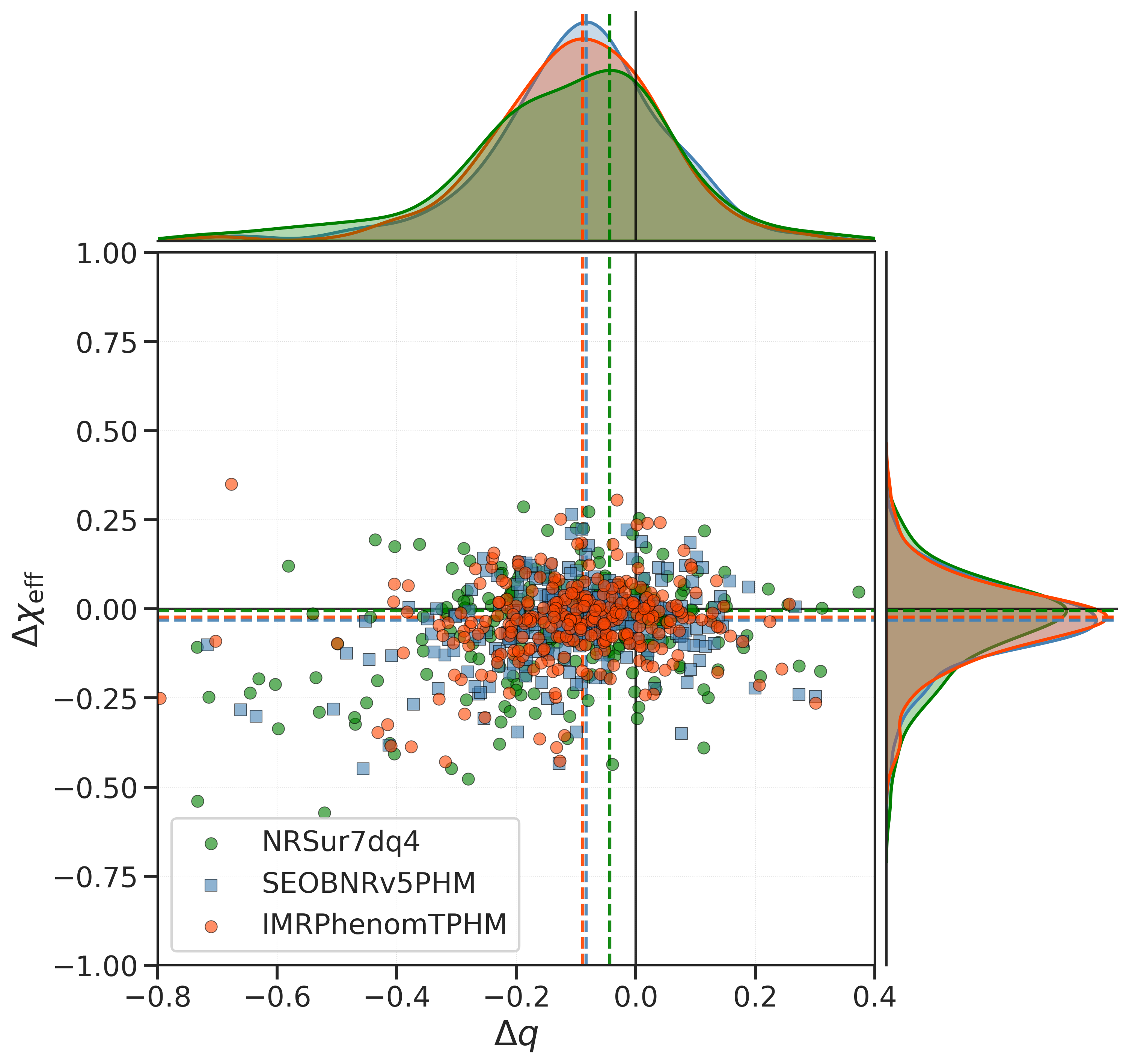}   \caption{\justifying{ Joint distribution of $\Delta\chi_\text{eff}$ vs $\Delta q$}. The marginal panels show the corresponding kernel density estimates, with dashed lines marking the peak of each distribution. The solid black lines mark zero shift.}
    \label{fig:delchi_vs_delq}
\end{figure*}

We now look at the joint distribution of the two shifts, $\Delta q$ and $\Delta\chi_{\rm eff}$, for all three recovery cases - \texttt{SEOBNRv5PHM}, \texttt{IMRPhenomTPHM}, and \texttt{NRSur7dq4} self-recovery - in Fig. \ref{fig:delchi_vs_delq}, and ask whether any correlation arises between the two parameter shifts. The marginal distributions confirm the individual parameter shrinkage seen in Sec. \ref{subsec:individual}: the peak of the $\Delta q$ distribution sits at $-0.083, -0.089, \text{ and } -0.049$ for SEOBNRv5PHM, IMRPhenomTPHM, and NRSur7dq4 self-recovery respectively and that for the $\Delta\chi_{\rm eff}$ distribution peaks at $-0.031, -0.023, \text{ and } -0.005$ for the same three cases. The bulk of the recovered population is offset from the true zero-shift centre for all three waveforms, indicating that part of the shift reflects ordinary PE behaviour due to noise and the excess beyond the self-recovery baseline attributable to inter-model systematics. We quantify this using the Kullback-Leibler (KL) divergence between the $\Delta q$ and $\Delta \chi_{\rm eff}$ distributions of each recovery case and the self-recovery case: $D_{\rm KL}(\Delta q) = 0.0431$ and $0.0596\,$ nats, and $D_{\rm KL}(\Delta\chi_{\rm eff}) = 0.0700$ and $0.0766\,$nats, for SEOBNRv5PHM and IMRPhenomTPHM respectively.

The cloud itself shows no obvious diagonal elongation, motivating a quantitative check. We quantify the coupling using both the Pearson correlation coefficient $r$, which measures the strength of a linear relationship, and the Spearman rank correlation $\rho$, which captures monotonic but non-linear relationships and is less sensitive to outliers. Reporting both allows us to confirm that any measured correlation is not an artefact of any extreme outliers. Table \ref{tab:correlations_combined} summarises the results. Recovery induces a modest but statistically significant positive correlation between $\Delta q$ and $\Delta\chi_{\rm eff}$ for both waveform families. All six $p$-values fall below the conventional $0.05$ significance threshold and the close agreement between the Pearson and Spearman values confirms this result is not driven by the handful of outliers. 

\begin{table}[htbp]
\centering
\caption{\justifying{Correlation coefficients between $q$ and $\chi_{\rm eff}$, and between their per-event shifts $\Delta q$ and $\Delta\chi_{\rm eff}$.}}
\label{tab:correlations_combined}
\renewcommand{\arraystretch}{1.4}
\begin{tabular}{lcc}
\hline
\hline
& Pearson $r$ ($p$-value) & Spearman $\rho$ ($p$-value) \\
\hline
\multicolumn{3}{l}{\textit{Correlation between $q_\text{rec}$ and $\chi_{\rm eff, rec}$}} \\
\hline
Injected (True)     & $-0.005$ ($0.93$) & $+0.003$ ($0.96$) \\
SEOBNRv5PHM  & $-0.083$ ($0.17$) & $-0.094$ ($0.12$) \\
IMRPhenomTPHM & $-0.062$ ($0.31$) & $-0.065$ ($0.28$) \\
NRSur7dq4 & $-0.088$ ($0.15$) & $-0.051$ ($0.40$) \\

\hline
\multicolumn{3}{l}{\textit{Correlation between $\Delta q$ and $\Delta\chi_{\rm eff}$}} \\
\hline
SEOBNRv5PHM         & $+0.253$ ($2.4\times10^{-5}$) & $+0.195$ ($1.2\times10^{-3}$) \\
IMRPhenomTPHM       & $+0.149$ ($1.4\times10^{-2}$) & $+0.144$ ($1.8\times10^{-2}$) \\
NRSur7dq4       & $+0.253$ ($2.4\times10^{-2}$) & $+0.144$ ($1.7\times10^{-2}$) \\
\hline
\hline
\end{tabular}
\end{table}

We also measured these values for the recovered parameters themselves. The injected catalog shows no correlation by construction ($r=-0.005$, $p=0.93$); however, all three recovered cases show a weak correlation of the same sign as the anti-correlation reported in GWTC catalogs, though not at a significant statistical significance ($p=0.12-0.40$ across all cases and both correlation measures). This is a reassuring result because the \texttt{NRSur7dq4} self-recovery case has zero waveform mismatch and yet shows almost the same non-significant recovered-value correlation as the other two cases, finding no evidence that waveform choices induce an excess correlation beyond ordinary PE.

A pattern that recurs across every result is that \texttt{SEOBNRv5PHM} and \texttt{IMRPhenomTPHM} agree closely with each other despite the two families being derived from structurally different approximation schemes (effective-one-body vs phenomenological). This agreement is quantified by the KL divergence between their $\Delta q$ and $\Delta \chi_{\rm eff}$ distributions: $D_{\rm KL}(\Delta q)=0.0087$ and $0.0081\,$ nats, and $D_{\rm KL}(\Delta\chi_{\rm eff})=0.0098$ and $0.0182\,$ nats taking \texttt{SEOBNRv5PHM} and $\texttt{IMRPhenomTPHM}$ as the reference distributions, respectively. The consistency means that the shifts we report are a shared limitation of current waveform modelling.

\subsection{Susceptibility of events in GWTC-5.0}
\label{subsec:gwtc5}
\begin{figure*}[htbp]
    \centering
  \includegraphics[width=\linewidth]{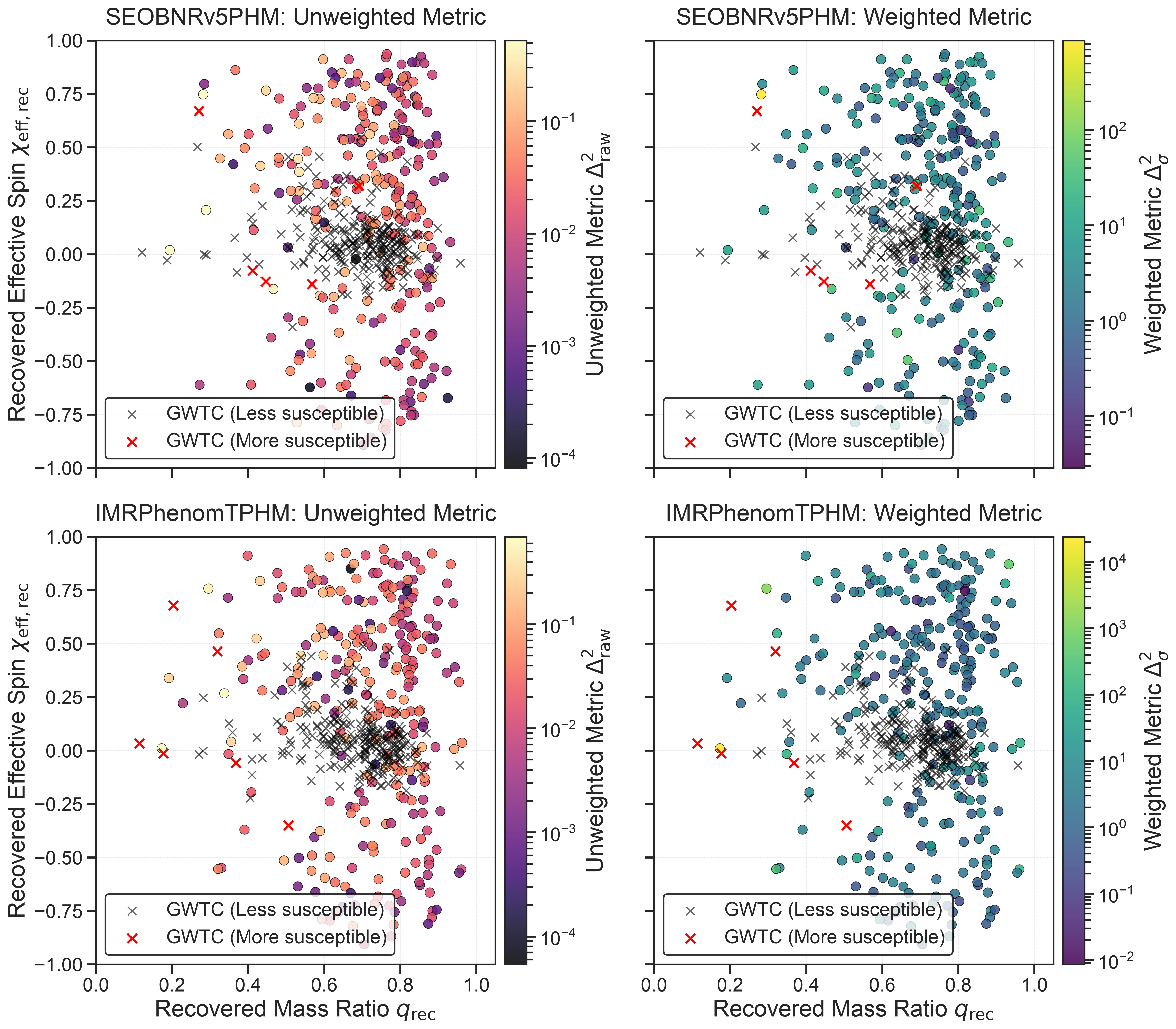}   \caption{\justifying{Recovered mass ratio $q_\text{rec}$ versus recovered effective spin $\chi_\text{eff,rec}$} for our injection catalog (circles, coloured by the unweighted metric $\Delta_\text{raw}^2$, left column, and the weighted metric $\Delta_\sigma^2$, right column as defined in Sec. \ref{subsec:biasdefn}, recovered using \texttt{SEOBNRv5PHM} (top row) and \texttt{IMRPhenomTPHM} (bottom row). Grey crosses mark the reported median for GWTC-5 events, with red crosses marking the events near a region where simulated injection(s) exceed a joint $3\sigma$ deviation in at least one recovery waveform. Some events exceed the threshold under both waveforms, while others exceed it under only one of the two.}
    \label{fig:gwtc5_overlay}
\end{figure*}

To assess whether the regions in $(q,\chi_{\rm eff})$ space identified as most susceptible to systematics in the previous section overlap with real, observed event source parameters, we compare our recovery against the GWTC-5.0 catalog \cite{GWTC5-results, GWTC5}, using posterior medians for the $267$ sources satisfying a false alarm rate (FAR) threshold of $ 1\, \text{yr}^{-1}$. Figure \ref{fig:gwtc5_overlay} overlays these medians directly onto the recovered-parameter plane, letting us ask whether currently detected events sit in the high-risk regions, and thus whether those events' inferred $(q, \chi_{\rm eff})$ values should be trusted at face value. For each recovery waveform, we use the GWTC posterior samples obtained with a model from the same family: \texttt{SEOBNRv4PHM/SEOBNRv5PHM} samples (for O3/O4 events) for comparison with our \texttt{SEOBNRv5PHM} recoveries, and \texttt{IMRPhenomXPHM} samples for comparison with our \texttt{IMRPhenomTPHM} recoveries.

The GWTC events densely cluster around $q\approx 0.6-0.9$ and low $\chi_{\rm eff} \approx -0.25 \text{ to }0.25$. No real events are recovered near median value $q \approx 1$, a distinct region of the GWTC population also identified by \cite{Region4}. Although this may also reflect normal PE behaviour due to railing against the prior, rather than waveform systematics. As a result, the weighted metric shows a small value, indicating that there is no bias in the inference. The comparison nonetheless gives concrete implications for how $(q,\chi_{\rm eff})$ structure in the current and future catalogs should be interpreted:

\begin{itemize}
    \item Individual events landing in high-displacement regions (Fig. \ref{fig:metric_colour}) should be treated with additional waveform-robustness checks before being cited as clear examples of a specific channel, say high-$\chi_{\rm eff}$, unequal-q hierarchical-merger candidates.

    \item The recovered values for asymmetric binaries with mass-ratios $q\lesssim0.6$ are more susceptible to systematic shifts from the true parameters, irrespective of the spin values. To quantitatively find events which lie in susceptible regions, we flag the ones where nearby simulated injections exceed a joint $3 \sigma$ deviation ($\Delta_{\sigma}^2 > 18$) in both waveforms. Two events stand out: GW190403$\_$051519 \cite{190403} and GW240527$\_$230910 \cite{GWTC5-results}, both of which lie within high-displacement regions in both recovery waveforms. These are low mass ratio systems, with GW190403$\_$051519 reported to have high positive effective spin and GW240527$\_$230910 having negative $\chi_{\rm eff}$. A further seven events exceed the threshold under only one of the two recovery waveforms: GW231113\_122623, GW240920\_073424 and GW241127\_061008 exceed the threshold under \texttt{SEOBNRv5PHM} only, while  GW200210\_092254, GW230518\_125908, GW241011\_233834 and GW241110\_124123 exceed it under \texttt{IMRPhenomTPHM} only. Events like these warrant careful waveform systematics check.
    

    \item The absence of significant shift in the joint $\chi_{\rm eff}-q$ distribution and any artificially induced correlation due to waveform uncertainty at O5 (A+) sensitivity shows the validity of the results obtained in GWTC-5.0, and the possible conclusions which can be drawn on sub-populations and formation channels drawn based on the correlation between the source parameters \cite{Transitions2026, Antonini2026}. 
    
\end{itemize}





\section{Discussion and Conclusion}
\label{sec:Disc_and_concl}

In this paper, we investigated to what extent waveform systematics can generate misleading structure in the recovered $\chi_{\rm eff}$-$q$ plane. We constructed a mock BBH catalog with $\chi_{\rm eff}$ drawn uniformly and independently of mass and compared how recovery using \texttt{SEOBNRv5PHM} and \texttt{IMRPhenomTPHM}, against an \texttt{NRSur7dq4} fiducial signal, introduces shifts in this signal. Unlike the mass-dependent shift identified in Paper-I \cite{Paper1}, which arises from the loss of inspiral cycles at higher masses, the shift reported here is driven by the intrinsic $\chi_{\rm eff}-q$ degeneracy in the PN phase (Sec. \ref{sec:Introduction}). We find that both $\chi_{\rm eff}$ and $q$ are pulled toward an intermediate recovered value. While the shifts in the two parameters are correlated, we found no statistically reliable correlation between the recovered values in our mock catalog. It is reassuring that at current waveform accuracy and detector sensitivity, this systematic does not manufacture a correlation, suggesting that existing astrophysical interpretations built on that correlation are not undermined. Our comparison with GWTC-5.0 data identifies regions of potential sensitivity in the parameter space of $q$ and $\chi_{\rm eff}$ which can be most impacted. We find that majority of the recovered parameter space is not strongly impacted by individual event-level waveform uncertainty to a degree that can impact any conclusion drawn on the formation channel. A small number of events, however, fall close to regions of maximum deviation in the $\chi_{\rm eff}$ and $q$ plane. GW190403$\_$051519 and GW240527$\_$230910 are identified under both recovery waveforms, while other events are identified under only one of the two waveforms. This study shows for the first time the reliability of waveform systematics in the effective spin and mass-ratio plane and demonstrates the possible parameter space of contamination that is important to identify for precision inference of astrophysical populations from GW data in the future. 

Several assumptions should also be kept in mind when interpreting our results. First, we adopt \texttt{NRSur7dq4} as a benchmark model to compare recovery against, rather than the absolute error relative to the physical signal. It is itself an approximate model, valid only for mass ratio $q\gtrsim1/6$ and spin magnitude $\lesssim0.8$ over approximately $20$ GW cycles \cite{NRSur_new}. Secondly, our injected population is restricted to spins aligned or anti-aligned with the orbital angular momentum. Waveform systematics are expected to be susceptible to the inclusion of spin precession since it introduces additional modelling complexity and further parameter degeneracies (e.g. between $\chi_{\rm p}$ and orbital inclination) on top of the $\chi_{\rm eff}$-$q$ degeneracy studied here \cite{spin_precession_th, spin_precession}. The shifts we report here are therefore a conservative lower bound on the true systematic effect for a generic, precessing BBH system. We have also not considered orbital eccentricity, which again introduces additional degeneracies with mass ratio and spin \cite{ecc_disc}. Thus, the framework that we developed in Paper-I \cite{Paper1} and implemented in this work is also extendable to a full precessing-spin catalog and to incorporate orbital eccentricity into the injected population.


\section*{Acknowledgments}
The authors are thankful to Aniruddha Chakraborty for carefully reviewing the manuscript as part of the LSC publication and presentation policy and providing useful feedback. This work is a part of the ⟨Data$|$Theory⟩ Universe Lab, which is supported by the Department of Astronomy and Astrophysics at the Tata Institute of Fundamental Research (TIFR) and the Department of Atomic Energy, Government of India. We acknowledge the support of the Department of Atomic Energy, Government of India, under Project Identification No. RTI 4012. This research is supported by the Prime Minister Early Career Research Award, Anusandhan National Research Foundation, Government of India. We are also thankful for the computing resources provided by the ⟨Data$|$Theory⟩ Universe Lab. LIGO, funded by the U.S. National Science Foundation (NSF), and Virgo, supported by the French CNRS, Italian INFN, and Dutch Nikhef, along with contributions from Polish and Hungarian institutes. This collaborative effort is backed by the NSF’s LIGO Laboratory, a major facility fully funded by the National Science Foundation. We acknowledge the use of the following software packages in this work: \texttt{NumPy \cite{numpy}, SciPy \cite{scipy}, Pandas \cite{pandas}, Bilby \cite{bilby_paper}, GWSim \cite{GWSim}, LALSuite \cite{lalsuite}, Matplotlib \cite{matplotlib}, and AstroPy \cite{astropy}.}

\appendix

\section{Mock Population Setup Details} \label{app:Setup deets}

The primary and secondary mass distributions are defined as 
\begin{widetext}
\begin{equation*}
\begin{aligned}
P_1\left(m_1 \mid m_{\min }, m_{\max }, \alpha\right) &= \begin{cases}(1-\lambda_g) \mathcal{P}\left(m_1,-\alpha\right)+\lambda_g G\left(m_1, \mu_g, \sigma_g\right), \quad \text{for } m_{\min }<m_1<m_{\max }, \\
0, \quad \text { otherwise }.\end{cases} \\
P_2\left(m_2 \mid m_{\min }, m_1, \beta\right) &= \begin{cases}\mathcal{P}\left(m_2,-\beta\right), \quad \text{for } \quad m_{\min }<m_2\leq m_1, \\
0, \quad \text { otherwise }.\end{cases}
\end{aligned}
\end{equation*}
\end{widetext}
Here, $\mathcal{P}\left(m_1,-\alpha\right)$ and $\mathcal{P}\left(m_2,-\beta\right)$ denote normalised power-law distributions with spectral indices $\alpha$ and $\beta$ respectively, defined over the specified mass ranges, and $m_{\min}$ is smoothed with a taper of width $\delta_m$ to avoid a sharp cutoff. The term $G\left(m_1, \mu_g, \sigma_g\right)$ represents a Gaussian probability density function with mean $\mu_g \text{ and standard deviation } \sigma_g$. The parameter $\lambda_g \in [0,1]$ represents the Gaussian mixing fraction between the power law and Gaussian components. The true hyperparameters used to generate the mock catalog are listed in Table \ref{tab:true_hyperparams}.

Effective spins $\chi_{\rm eff}$ are drawn independently from a uniform distribution $\mathcal{U}(-1,1)$. Spin orientations are isotropic: tilt angles $\theta_1$ and $\theta_2$ (between each spin and the orbital angular momentum) are drawn uniformly in $\cos\theta \in [-1, 1]$ and azimuthal angles $\phi_{12}$ and $\phi_{JL}$ are drawn uniformly in $[0, 2\pi)$. The detector selection effects are incorporated using the O5 noise PSD, and an SNR cutoff is set at 12. The resulting detected events constitute the mock catalog used in subsequent analyses. The priors on the PE parameters of masses and spins are listed in Table \ref{tab:priors}. Spin tilt angles $\theta_1 \text{ and }\theta_2$, azimuthal angles $\phi_{12} \text{ and } \phi_{JL}$ and luminosity distance $d_L$ are fixed at their injected values. 

\begin{table}[htbp]
\centering
\caption{\justifying True population hyperparameters used to generate the  Power Law+Gaussian
mock catalog.}
\label{tab:true_hyperparams}
\begin{tabular}{lc}
\hline
Parameter & True injected value \\
\hline
$\alpha$     & 3.5 \\
$\beta$      & 1.1  \\
$m_{\min}\,[M_\odot]$  & 5     \\
$m_{\max}\,[M_\odot]$  & 100  \\
$\delta_m\,[M_\odot]$  & 4.8    \\
$\mu_g\,[M_\odot]$     & 34  \\
$\sigma_g\,[M_\odot]$  & 4   \\
$\lambda_g$  & 0.038  \\
\hline
\end{tabular}
\end{table}


\begin{table}[t]
\centering
\caption{\justifying Priors on  event-level PE parameters used in the injection-recovery analysis. }
\label{tab:priors}
\begin{tabular}{lc}
\hline
Parameter & Prior \\
\hline
$\mathcal{M}$ & $\mathcal{U}(3,170)$ \\
$q$ & $\mathcal{U}(0.1,1)$ \\
$a_1,\,a_2$ & $\mathcal{U}(0,0.99)$ \\

\hline
\end{tabular}
\end{table}

\bibliographystyle{apsrev4-1}
\bibliography{main_paper}

\end{document}